\documentclass[10pt]{article}

\usepackage[english]{babel}
\usepackage[T1]{fontenc}
\usepackage{epsfig}
\usepackage{multicol}
\usepackage{amsmath, amsthm}
\usepackage{amsfonts,amssymb}
\usepackage{float}
\usepackage{setspace}
\usepackage{url}
\usepackage{enumitem}
\usepackage{amstext}
\usepackage{xcolor}
\usepackage{amsfonts}
\usepackage{amssymb}
\usepackage{graphics}
\usepackage{graphicx}
\usepackage{pstricks,pst-node,pst-tree}
\usepackage{enumerate}
\usepackage{tikz} 
\usepackage{cutwin}
\usepackage{xcolor}
\usepackage{colortbl}
\usepackage{algorithm,algorithmic}
\usepackage{multirow}
\usepackage{amsthm}
\usepackage{glossaries}
\usepackage{array,arydshln}
\usetikzlibrary{shapes,matrix}
\usepackage{comment}
\usepackage{mathrsfs}
\usepackage{authblk}
\usepackage{wrapfig}
\usepackage[font=footnotesize]{caption}
\usepackage{stackengine}
\usepackage{caption}
\usepackage{subcaption}
\usepackage{hyperref}
\usepackage{stmaryrd}
\usepackage{bbm}
\usepackage{relsize}
\usepackage{pifont}
\usepackage{biblatex}
\usepackage{hyperref}
\usepackage[left=2cm,right=2cm,top=2cm,bottom=2cm]{geometry}
\usepackage{xr}
\makeatletter

\usepackage{biblatex}
\usepackage{hyperref}

\newcommand*{\addFileDependency}[1]{
\typeout{(#1)}
\@addtofilelist{#1}

\IfFileExists{#1}{}{\typeout{No file #1.}}
}\makeatother

\newcommand*{\myexternaldocument}[1]{
\externaldocument{#1}
\addFileDependency{#1.tex}
\addFileDependency{#1.aux}
}

\myexternaldocument{Annex}

 \title{A Boolean encoding of the Most Permissive semantics\\ for Boolean networks}
\author[1]{Laure de Chancel}
\author[3]{Brigitte Mossé}
\author[2]{Aurélien Naldi}
\author[3]{Elisabeth Remy}

\affil[1]{Institut Curie, Génétique et biologie du développement (UMR3215 / U934), Paris, France}
\affil[2]{IBM France, Bois-Colombes, France}
\affil[3]{Aix Marseille Univ, CNRS, I2M (UMR 7373), Turing Center for Living systems, Marseille, France}

\begin{document}

\maketitle
\vspace{1cm}
\paragraph{Abstract} 

Boolean networks are widely used to model biological regulatory networks and study their dynamics. Classical semantics, such as the asynchronous semantics, do not always accurately capture transient or asymptotic behaviors observed in quantitative models. To address this limitation, the Most Permissive semantics was introduced by Paulevé et al., extending Boolean dynamics with intermediate activity levels that allow components to transiently activate or inhibit their targets during transitions. In this work, we provide a Boolean encoding of the Most Permissive semantics: each component of the original network is represented by a triplet of Boolean variables, and we derive the extended logical function governing the resulting network. We prove that the asynchronous dynamics of the encoded network exactly reproduces the attainability properties of the original network under Most Permissive semantics. This encoding is implemented as a modifier within the bioLQM framework, making it directly compatible with existing tools such as GINsim. To address scalability limitations, we further extend the tool to support partial unfolding, restricted to a user-defined subset of components. 
\newpage
\tableofcontents

\vspace{1cm}

\section{Introduction}
Boolean networks have established themselves as a powerful formalism for modeling biological regulatory networks, including gene regulatory networks and cell signaling pathways. In this framework, each component of the system — a gene, a protein, or a signaling molecule — is represented by a binary variable, and the dynamics of the network is governed by a set of logical functions encoding the regulatory interactions between components. The simplicity and interpretability of this formalism have made it a method of choice for studying the qualitative behavior of complex biological systems, including the identification of stable states and reachable configurations that reflect the possible phenotypes of a cell.
The dynamics of a Boolean network is determined not only by its logical functions but also by the chosen updating semantics, which specifies how and in which order the components are updated. Two semantics are classically used: the synchronous semantics, in which all components are updated simultaneously, and the asynchronous semantics, in which a single component is updated at a time, leading to a non-deterministic dynamics. While these semantics are convenient and well-studied, they sometimes fail to capture transient behaviors observed in quantitative models - in particular, the transient activation of a component that never reaches a sustained active state but nonetheless influences the downstream network during the transition.
To overcome this limitation, Paulevé et al. introduced the Most Permissive semantics, in which each component can occupy, in addition to its Boolean active and inactive levels, two intermediate levels corresponding to an increasing or a decreasing transition. Components at these intermediate levels are treated as simultaneously active and inactive by their targets, allowing the Most Permissive set of downstream transitions to be explored. This semantics provably subsumes all other updating semantics — any trajectory reachable under any semantics is also reachable under Most Permissive — making it particularly suited for reachability analysis and for reconciling Boolean predictions with quantitative observations.
However, the Most Permissive semantics operates over an extended state space that is not directly supported by most existing tools for Boolean networks, which are designed around classical Boolean dynamics. This creates a practical gap between the theoretical expressiveness of the Most Permissive semantics and its accessibility within standard modeling pipelines.
In this article, we close this gap by providing a Boolean encoding of the Most Permissive semantics. Our construction maps each component of the original n-nodes network to a triplet of Boolean variables, yielding a 3n-node Boolean network whose asynchronous dynamics exactly reproduces the reachability properties of the original network under Most Permissive semantics. We prove the correctness of this encoding and implement it as a modifier in the bioLQM framework, making the encoded model directly compatible with tools such as GINsim. To address the scalability challenges inherent to the tripling of the number of variables, we additionally implement a partial unfolding variant, in which only a chosen subset of components is extended, preserving reachability analysis capabilities while limiting model inflation. We illustrate our approach on a three-component example and on a published 15-component model of hematopoiesis, and discuss the structure of the extended regulatory graph and its relationship to the original network.

The article is organized as follows. Section 1 recalls the necessary background on Boolean networks, classical semantics, and the Most Permissive semantics. Section 2 presents the Boolean encoding and the construction of the extended logical function. Section 3 describes the implementation within bioLQM and discusses the results, tests, and limitations of the approach. Section 4 concludes with perspectives for future work.
Mathematical details and proofs are given in \textsc{appendix}.

\subsection{Boolean networks}

We consider a system constituted of $n$ components.
We associate to this system a Boolean network defined by a logical function $f=(f_1, \dots, f_n): \mathbb{B}^n \xrightarrow{} \mathbb{B}^n$, with $\mathbb{B} = \{0,1\}$, and a semantics that specifies the evolution of the binary vectors (elements $x= (x_1, \dots, x_n)$ of $\mathbb{B}^n$).
For each $j=1,\dots n$, the coordinate $x_j$ represents the level of the $j$th component ($1$ active, $0$ inactive). Binary vectors are also called configurations or states of the system. 
A configuration $x$ is said to be stable under $f$, or a fixed point of $f$, if $f(x)=x$.

\begin{figure}[h]
\begin{minipage}{8cm}
    $f=(f_1, f_2, f_3): \mathbb{B}^3 \xrightarrow{} \mathbb{B}^3 \\
    \\
    f_1(x) = x_1 \land \neg x_3 \\
    f_2(x) = x_1\\
    f_3(x) = \neg x_1
    $
   \end{minipage}
   \begin{minipage}{5cm}

    \begin{center}
            \includegraphics[width=7cm]{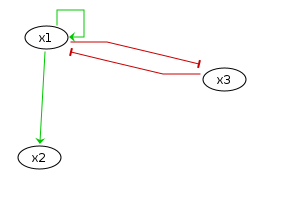}
    \end{center}
       \end{minipage}
\caption{\label{reg_a}Example A. Left: logical rules. Right: associated regulatory graph.} 
 \end{figure}

Logical functions are expressed with operators $\scriptstyle{AND}$ ($\land$), $\scriptstyle{OR}$ ($\lor$) and $\scriptstyle{NOT}$ ($\neg$). These rules reflect the regulations between components of the system, possibly self-regulations. Thus, we associate to the Boolean network a regulatory graph, that is a signed directed graph whose nodes represent the components, and directed edges represent the regulations between components (see Example A Figure \ref{reg_a}). The sign of the edges reflect the nature of the regulations (positive edges for activations, negative for inhibitions). The sign can vary depending on the context (edges with undetermined signed). 
Note that the regulatory graph is deduced from the logical functions, but does not contain all the information of the Boolean network: it does not allow to know whether we are dealing with conjunctions or disjunctions between the regulations.

\bigskip

\begin{figure}[H]
    \begin{minipage}[c]{.12\linewidth}
        \begin{center}
            \includegraphics[width=1cm]{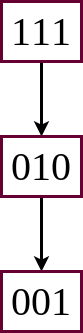}
            \title{Synchronous}
            \end{center} 
            \hfill
    \end{minipage}
    \begin{minipage}[c]{.4\linewidth}
        \begin{center}
            \includegraphics[width=4.6cm]{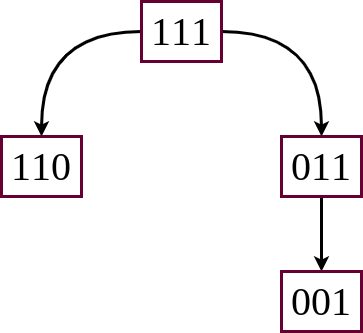}
            \title{Asynchronous}
            \end{center} 
            \hfill
    \end{minipage}
    \begin{minipage}[c]{.4\linewidth}
            \begin{center}
            \includegraphics[width=5cm]{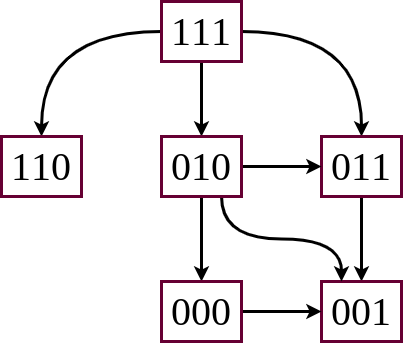}
            \title{Generalized asynchronous semantics}
            \end{center}
            \hfill
    \end{minipage}
    \caption{State transition graphs from configuration $111$ for Example A. Left: synchronous STG. Middle: asynchronous STG. Right: Generalized asynchronous STG.}
    \label{stg_A}
 \end{figure}

Whereas the logical function $f$ specifies, for each configuration, the level towards which each component tends, the semantics defines the evolution of the system, i.e. the succession of configurations  \cite{pauleve}. 
Two semantics are mainly used in the literature. The synchronous semantics updates the level of all the components simultaneously, following the values specified by $f$ \cite{KAUFFMAN1969437}, leading to a deterministic behavior. The asynchronous semantics updates the level of a single component at a time \cite{THOMAS1973563}. Hence, each configuration has as many successors as the number of components called to update by the logical function $f$ (the number of $i$ such that $f_i(x)\neq x_i$), leading to a non-deterministic behavior. A variant of the latter is the generalized asynchronous semantics: each subset of components is allowed to update synchronously at a time.
The trajectories of the Boolean network are represented with a state transition graph (STG), whose nodes are the configurations of the network, and the edges represent the possibility of moving from one configuration to another. The terminal strongly connected components of the STG are called the attractors. An attractor is simple when reduced to a single configuration (stable state), complex otherwise. The attractors represent the asymptotic behaviors of the system.
At least for small values of $n$, the STG enables to visualize the stable configurations and reachable properties that are the existence of a path between two sets of configurations. Figure \ref{stg_A} represents the STGs of Example A for the different semantics.

\subsection{Most Permissive semantics  } \label{ss:def-m.p.}

The previous classic semantics allow us to get a wide variety of dynamics. However, there are situations where quantitative studies or biological observations highlight trajectories that are impossible to obtain with these Boolean models. In \cite{modele_b}, this is illustrated through an example 
in which these semantics do not allow for the transient activation of one of the components, which can be observed in quantitative studies. This observation led to the introduction of Most Permissive semantics. 

\begin{figure}[H]
    \begin{minipage}[c]{7.5cm}
        \begin{center}
            \includegraphics[width=7cm]{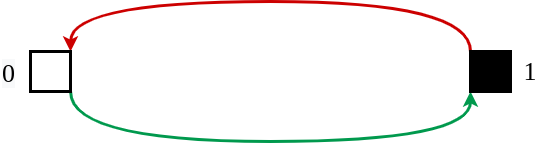}
            \title{(a)}
            \end{center} 
            \hfill
    \end{minipage}
    \begin{minipage}[c]{7.5cm}
        \begin{center}
            \includegraphics[width=7cm]{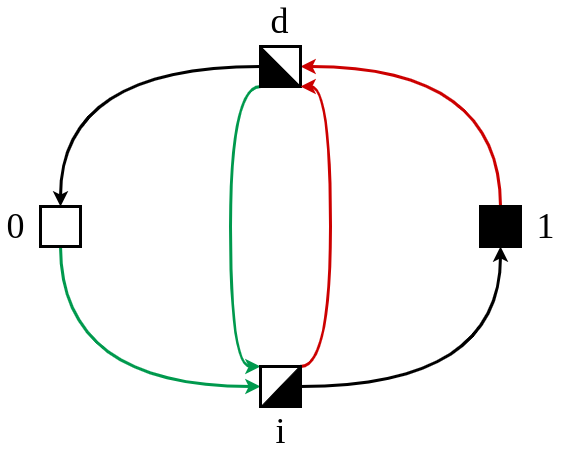}
            \title{(b)}
            \end{center} 
    \end{minipage}
    \centering
    \caption{Left: Schema of transitions between levels for a Boolean variable. Right: Schema of transitions between levels for a variable in Most Permissive semantics. Black arrows indicate a guaranteed transition, green arrows indicate a transition under positive control, and red arrows indicate a transition under negative control.}
    \label{bool_vs_perm}
\end{figure}

In Most Permissive semantics, for each component, two intermediate levels are introduced in addition to the two Boolean levels $0$ and $1$ (cf Figure \ref{bool_vs_perm}): level $i$ for \textit{increasing} (between inactive level $0$ and active level $1$), and level $d$ for \textit{decreasing} (between active level $1$ and inactive level $0$). 
A Boolean configuration is one with all components at Boolean levels.

The special feature of Most Permissive semantics is that the two intermediate levels can be seen as both active and inactive levels by the components of the Boolean network.  This means that when a component is at level $i$ or $d$, its targets will consider the level that as far as possible makes their regulation effective.

Instead of the two transitions controlled by regulators of the Boolean network in Figure \ref{bool_vs_perm} (Left), we have six transitions in Figure \ref{bool_vs_perm} Right: four of them are controlled by regulators and two are guaranteed (always possible and not dependent on the rest of the network). Note that, even if the components are regulated in two steps (from level $0$ to level $1$, or from level $1$ to level $0$), only one step is controlled (see Figure \ref{bool_vs_perm} green and red edges).

We now define the Most Permissive semantics more formally\label{formalMP}. We consider a Boolean network $f : \mathbb{B}^n \xrightarrow[]{}\mathbb{B}^n,\\ f=(f_1, ... f_n)$ and the set $X_{m.p.} = \{ 0, i, d, 1\}^n$, where $i$ and $d$ are abstract levels.

For $x=(x_1, \dots , x_n) \in X_{m.p.}$, let us consider the set 
$$\gamma(x) = \{ x'=(x_1', ... x_n') \in \mathbb{B}^n\;;\; \forall i \in \{1, ..., n\}, \;x_i \in \{0,1\} \implies x_i' = x_i \}\;.$$

\noindent The (asynchronous) Most Permissive semantics associated with $f$ is then given as follows: \\
Let $x,y \in X_{m.p.}$; there is a transition from $x$ to $y$ if there exists a unique $i_0 \in \{1, \dots,n\}$ such that $x_{i_0} \neq y_{i_0}$, and one of the following situations occurs:

\begin{itemize}[label=$-$] 
   
   \item ($x_{i_0} = 0$ or $d$) and ($\exists x' \in \gamma(x)$ s.t. $f_{i_0}(x') = 1$) and ($y_{i_0} = i $),
   \item ($x_{i_0} = 1$ or $i$) and ($\exists x' \in \gamma(x)$ s.t. $f_{i_0}(x') = 0$) and ($y_{i_0} = d$),
    \item ($x_{i_0} = i$) and ($y_{i_0} = 1 $),
    \item ($x_{i_0} = d$) and ($y_{i_0} = 0 $). 
\end{itemize}

\paragraph{Remark}
Note that this dynamics is non-deterministic at two degrees: there can be several successors to $x$ relative to different values of $i_0$, but also for a given value of $i_0$ relative to different values of $y_{i_0}$. Denoting $f_{m.p.}$ the synchronous function associated with the Most Permissive semantics, we thus have $f_{m.p.} : \{0, i, d, 1\}^n \xrightarrow{} \mathcal{P} \{0, i, d, 1\}^n $, with $\mathcal{P} \{0, i, d, 1\}^n$ the set of parts of $\{0, i, d, 1\}^n$. For instance, for Example A, if we start from state $i00$, the first component can change in two different ways, giving either 1 or $d$, so that $100$ and $d00$ are both successors of $i00$ in the Most Permissive semantics.

\begin{figure}[H]
    \centering
    \includegraphics[width = 7cm]{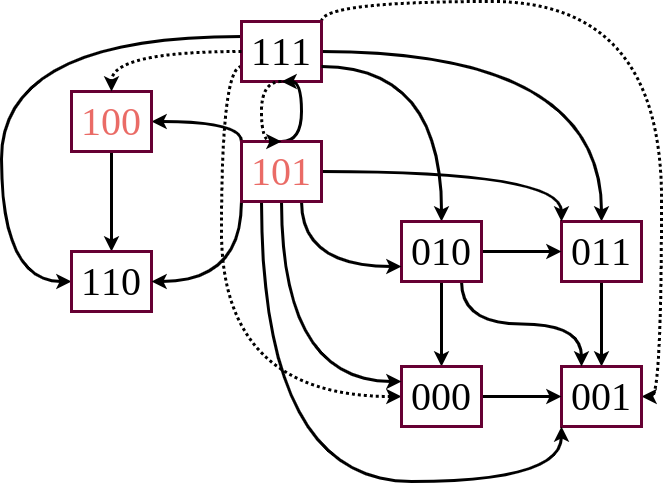}
    \caption{STG from initial configuration $111$ for Example A in Most Permissive semantics. Only Boolean configurations are shown. Configurations in red are those reachable with Most Permissive semantics, but not with classical semantics. Solid arrows represent transitions that are possible with classical semantics, while dotted arrows are those that pass through the increasing or decreasing level of Most Permissive. }
    \label{mp_a}
\end{figure}

\medskip
To visualize the trajectories of Most Permissive semantics and compare them with classical semantics, we can perform STGs restricted to regular configurations. This STG for Example A is shown in Figure \ref{mp_a}. We note that configurations not reachable in asynchronous and synchronous semantics become reachable, and as expected stable configurations remain the same (cf Figure \ref{stg_A}). The set of all Most Permissive transitions for Example A is given in \textsc{appendix} 0 Table \ref{MP exA}.\\

\subsection{Modeling tools}

We have several tools for manipulating Boolean networks, developed by CoLoMoto (Consortium for Logical Models and Tools). Some of these tools are of particular interest to us: 
\begin{itemize}[label=$-$] 
    \item BioLQM is a Java tool for converting, transforming and analyzing logical models \cite{biolqm};
    \item GINsim allows to model and run simulations of genetic regulation networks \cite{ginsim}; it is a graphical interface based on BioLQM.  
\end{itemize}
These tools can be used to create Boolean networks and convert them into different formats. We can compute stable configurations, choose between different semantics, and simulate trajectories based on one or more configurations.

\section{Boolean encoding of the Most Permissive semantics}

Our objective is to properly encode the Most Permissive dynamics (represented Figure \ref{bool_vs_perm} Right) by an asynchronous Boolean dynamics. This Boolean mapping will give access to a set of existing tools, in particular GINsim, and thus will enable to study the Most Permissive dynamics. In this section, we give a detailed description of the encoding, and how to obtain the associated extended Boolean function. In \textsc{appendix} 1, we prove the validity of the encoded model: consistent definition of a Boolean function which translates the Most Permissive dynamics, in the sens that it reproduces exactly the attainability between Most Permissive states.

\subsection{Encoding with three Boolean variables}
We represent the level of each component by a triplet of Boolean variables. Thus, to each Most Permissive configuration $x = (x_1,\dots, x_n)$, with $x_j \in \{0,1,i,d\}$, of a model with $n$ components is associated a configuration $x = (x_{1a}, x_{1b}, x_{1c},\dots,x_{na}, x_{nb}, x_{nc})$, with $x_{jk} \in \{0, 1\}$ for $1\leq j\leq n$ and $k \in \{a, b,c\}$. We choose to encode the Most Permissive levels $0$, $1$, $i$ and $d$ in the following way:
\begin{itemize} 
    \item the inactive level $0$ by the triplet $000$, 
    \item the active level $1$ by the triplet $111$,
    \item the level $i$ by the triplet $001$,
    \item the level $d$ by the triplet $101$.
 \end{itemize}   
Moreover,
\begin{itemize} 
    \item triplet $011$ is inserted between $001$ and $111$,
    \item triplet $100$ is inserted between $101$ and $000$,
    \item triplet $010$ is sent to $000$,
    \item triplet $110$ is sent to $111$.
     \end{itemize}  

\begin{figure}[H]
    \centering
    \includegraphics[width=8cm]{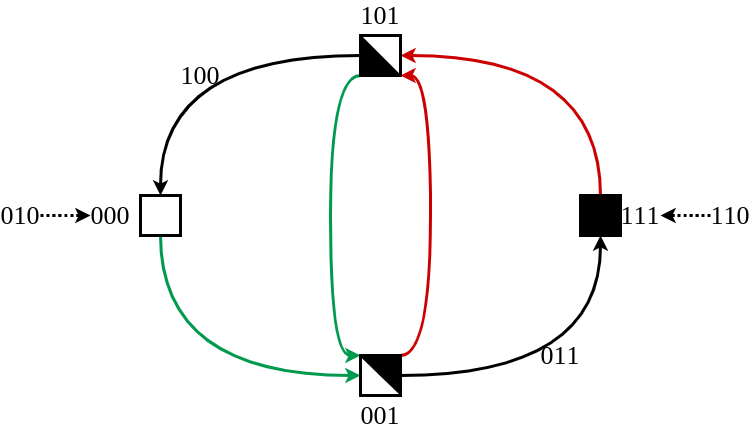}  
    \caption{Encoding with three Boolean variables of the Most Permissive semantics. Green and red edges represent controlled transitions, black edges are guaranteed transitions.}
    \label{Encodage}
\end{figure}

     The choices of the above encoding of $0$ and $1$ are the most natural ones. We have chosen to encode $i$ by $001$ (resp. $d$ by $101$) so that the controlled transition from $0$ to $i$ (resp. from $1$ to $d$) is asynchronous.
     Also, the intermediate triplets $011$ and $101$, inserted in  the guaranteed transitions from $001$ to $111$ and from $101$ to $000$, are required for maintaining the asynchronicity. 
     Finally, triplets $010$ and $110$ are sent to $000$ and $111$ respectively, in accordance with asynchronicity. We call them artifacts triplets (since they are not involved in the Most Permissive trajectories). Figure \ref{Encodage} represents this encoding. Note that within this encoding, both activation and inhibition of a component require three steps (instead of two in the original Most Permissive), and still only one step is controlled. 
    
\paragraph{Remark} 
Any two-variables encoding does not meet asynchronous constraints.

\subsection{The extended logical function} 
Given the logical function $f : \mathbb{B}^{n} \xrightarrow{} \mathbb{B}^{n}$, we have to specify the extended logical function $f_{ext} : \mathbb{B}^{3n} \xrightarrow{} \mathbb{B}^{3n}$ which governs the variables of the encoded Most Permissive scheme applied to $f$.

We first consider "internal rules" describing interactions between Boolean variables of the triplet corresponding to a component (that are independent of $f$), and then the regulation rules between triplets.

\subsubsection{Internal rules} \label{internal}
Internal rules are directly obtained from the definition of the Boolean encoding (cf Figure \ref{Encodage}). Considering the $j$th component, represented by the triplet of variables $(x_{ja}, x_{jb}, x_{jc})$, we specify for each $k\in\{a,b,c\}$ the values of this triplet such that $x_{jk}$ is updated to $1$ under $f_{ext}\,$.

Let $f_{ext}^{-1}(1**)_j$ denote the list of triplets $(x_{ja}, x_{jb}, x_{jc})$ for $x_{ja}$ to be updated to $1$.
Triplets $011$, $110$ and $111$ satisfy naturally this condition (see Figure \ref{Encodage}).
There are still two ways of setting the first variable to $1$. On the one hand, being at triplet $001$ with conditions of inhibition of the component effective (denoted by $001 \land \ominus_j$). On the other hand, being at triplet $101$ with conditions of activation of the component non effective ($101 \land \neg \oplus_j$).
We therefore obtain $\;\;f_{ext}^{-1}(1**)_j : 011, \;110, \;111, \;001 \land \ominus_j,\;101 \land \neg \oplus _j. $ 

\noindent In the same way, we have: 
\begin{align*}
&f_{ext}^{-1}(1 * *)_j : 011,\; 110,\; 111,\; 001 \land \ominus_j,\;101 \land \neg \oplus_j   \\
&f_{ext}^{-1}(* \, 1 \, *)_j :110,\; 0*1, \;111 \land \neg \ominus_j  \\
&f_{ext}^{-1}(*  *  1)_j :11*,\; 0*1,\; 000 \land \oplus_j  
\label{the internal rules}
\end{align*}
\noindent Note that we need the four notations $\oplus_j$,  $\neg \oplus_j$,  $\ominus_j$ and  $\neg \ominus_j$, due to the fact that $\oplus_j$ and $\ominus_j$ are non complementary. This will be illustrated in next section. 

\subsubsection{Regulation rules} \label{regulation}

A component is considered as active when its Most Permissive levels are \textit{i}, \textit{d} or $1$, that correspond to triplets $001$, $101$ and $111$. It is considered as inactive when its Most Permissive levels are \textit{i}, \textit{d} and $0$, that correspond to triplets $001$, $101$ and $000$.

Thus, active components are characterized by the third variable of their triplet equal to $1$ (the transient triplet $011$ is consistent). Inactive components are characterized by the second variable equal to $0$ (as the transient triplet $100$).
Artifacts triplets $110$ and $010$ are neither active nor inactive following these rules, as expected. To summarize:
\begin{itemize}[label=$-$] 
    \item components considered as active  correspond to triplets $(x_{ja}, x_{jb}, x_{jc})$ of $\ **1\ $
    \item components considered as non active  correspond to triplets of $\ **0$
    \item components considered as inactive  correspond to triplets of $\ *\,0\,*$
    \item components considered as non inactive correspond to triplets of $\ *\,1\,*\,\;$
\end{itemize}
\paragraph{Remark} The first variable of the triplets is not involved in the above characterizations. 
Note that the shortest path from $000$ to $111$ contains only states with  $x_{ja} = 0$. Also, the shortest path from $111$ to $000$ contains only states with  $x_{ja} = 1$.
\medskip

Logical functions $f_j$ are conditions written with conjunctions, disjunctions and negations of Boolean variables. 
It remains to make the corresponding conjunctions, disjunctions and negations to get the regulatory conditions $\oplus_{j}$, $\neg\oplus_{j}$, $\ominus_{j}$ and $\neg\ominus_{j}$. We then inject these conditions in the internal rules obtained in Section \ref{internal}.
This defines the logical function $f_{ext}$ which encodes the Most Permissive semantics, using the asynchronous semantics. 

\medskip
\noindent For instance, with $f_1(x) = x_1 \land  \neg x_3$ in Example A (cf Figure \ref{reg_a}), we get the following regulatory conditions: \\
$\oplus_{1} \;\;:\; **1 \ *** \ *0* $ \\
$\neg \oplus_{1} :\; **0 \ *** \ *** \lor *** \ *** \ *1* $ \\
$\ominus_{1} \;\;: \;*0* \ *** \ *** \lor *** \ *** \ **1 $\\
$\neg \ominus_{1} :\; *1* \ *** \ **0 $\\
Injecting this in the internal rule 
$f_{ext}^{-1}(1**)_1 : 011, \;110, \;111, \;001 \land \ominus_1,\;101 \land \neg \oplus _1$, we obtain:\\
$f_{ext}^{-1}(1**\ ***\ ***) = 011\ ***\ ***\;\cup\;110\ ***\ ***\;\cup\;111\ ***\ ***\;\cup\;001\ ***\ ***\;\cup\;101\ *** \ *1* $.\\

\begin{figure}[H]
    \centering
    \includegraphics[width=12cm]{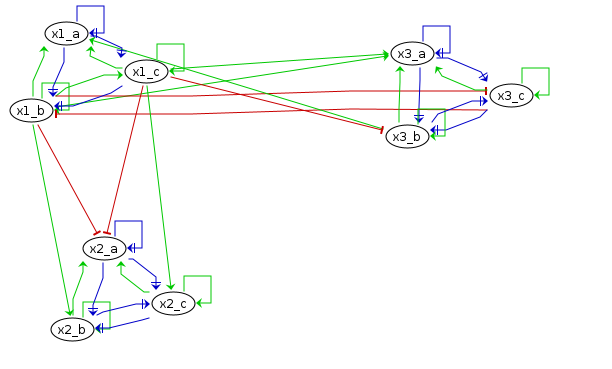}
    \caption{Regulatory graph associated with the extended Example A.
    Red arrows represent negative regulations, green arrows positive regulations. Blue arrows represent regulations with undetermined sign.}
    \label{etendu}
\end{figure}

This calculation has been performed for all variables of Example A (cf \textsc{appendix} 2); the resulting regulatory graph of the unfolded model is shown in Figure \ref{etendu}.

We can observe on Figure \ref{etendu} similar patterns between the variables of each triplet, as expected from internal rules. We give more details in \textsc{appendix} 3 for guaranteed regulations and non-guaranteed regulations between variables of a triplet.
The regulations between different components of the initial model are still visible: we can observe an inhibition of $x_{3c}$ by $x_{1b}$ and of $x_{3b}$ by $x_{1c}$, an inhibition of $x_{1b}$ by $x_{3c}$, an activation of $x_{2b}$ by $x_{1b}$ and of $x_{2c}$ by $x_{1c}$. However, we also find for instance an activation of $x_{1a}$ by $x_{3b}$. The fact that this regulation by $x_{3b}$ is positive, i.e. of opposite sign to the regulation of the initial model, accounts for the greater complexity of the extended model, compared with the initial one (see more details in \textsc{appendix} 4).

\section{Implementing Boolean unfolding}
\subsection{Creation of a bioLQM Modifier}
We have coded an extension to the bioLQM code that gives from a $n$-Boolean model a $3n$-Boolean extended model that can be interpreted as Most Permissive semantics.
In bioLQM, a \textit{Modifier} is a service that takes a model as input (sometimes with parameters) and returns a modified or constructed model from the initial model. We have instantiated a \textit{Modifier}, which takes a model as input and returns the extended model described above. This extended model is compatible with GINsim. 
The code produced is in Java, and uses bioLQM structures. These pre-existing structures provide a framework for coding without starting from scratch, and offer a good solution for managing complex objects, such as logical relationships between components. On the other hand, it requires adaptation to the constraints imposed by the framework, and a good understanding of a large amount of pre-existing code.

More precisely, we use the pre-existing \textit{Modifier} interface, which we extend. We therefore take as input a model in a compatible format (.bnet for instance) and return a model in this format.    
We then need to create a new model from the first, by extending the components and logical functions. The new components are obtained from the first by creating three variables for one of the initial model. Then we need to integrate the internal logical rules between the components of each triplet representing an extended component, and inject the logical rules between components. 
We use MDDs (Multi-valuated Decision Diagrams, here used as binary decision diagrams) to build and store logical relationships between components. An internal bioLQM library, MDDlib, is used to manipulate them. We need to be careful about memory allocation, and free up allocated space that we are no longer using as we go along.

\medskip
The code has been integrated into bioLQM in the package \textit{org.colomoto.bioloqm.modifier.most\_permissive} and is available at this address: 
\ \href{https://github.com/colomoto/bioLQM}
{https://github.com/colomoto/bioLQM}.\\

Using \textit{Modifier most\_permissive} requires bioLQM to be installed, instructions are available here:

\href{http://colomoto.org/biolqm/doc/install.html}{http://colomoto.org/biolqm/doc/install.html}.

\subsection{Tests and limitations}

We applied the method on the following three components model; we obtained reachability results consistent with those described in \cite{loic_roncalli}.
\begin{eqnarray*}
   && f=(f_1, f_2, f_3): \mathbb{B}^3 \xrightarrow{} \mathbb{B}^3 \\
   &&f_1(x) = signal \\
  &&  f_2(x) = x_1\\
  && f_3(x) = \neg x_1 \land x_2
\end{eqnarray*}

\begin{figure}[H]
    \centering
    \includegraphics[width = 10cm]{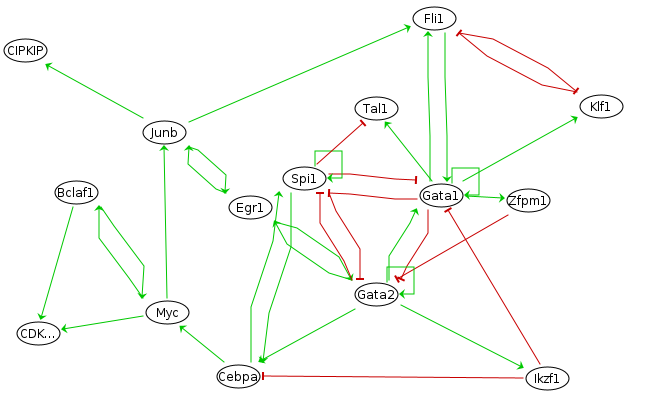}
    \caption{Regulatory graph of Example B.}
    \label{mod_c}
\end{figure}

The limiting factor of this method is clearly scaling. 
For instance, in a model encompassing $15$ components as in \cite{leonard} (cf Figure \ref{mod_c}), the extended model contains $15\times 3=45$ components. The space of configurations, and even the one of reachable configurations, become too large to be represented explicitely. Using GINsim, it is not possible to assess the reachability of configurations.

However, it may be desirable to unfold only a subset of the components. 
Chosen components are transformed into triplets with associated internal rules. The extended components are integrated into the network through their internal control functions, as described in Section \ref{regulation}. The use of the partial unfolding implies knowing which components are worth extending. 

This partial development is equivalent to partial MP dynamics, proposed and applied to the model of example B in \cite{benboina25a}.

\begin{figure}[H]
    \centering
    \includegraphics[width = 12cm]{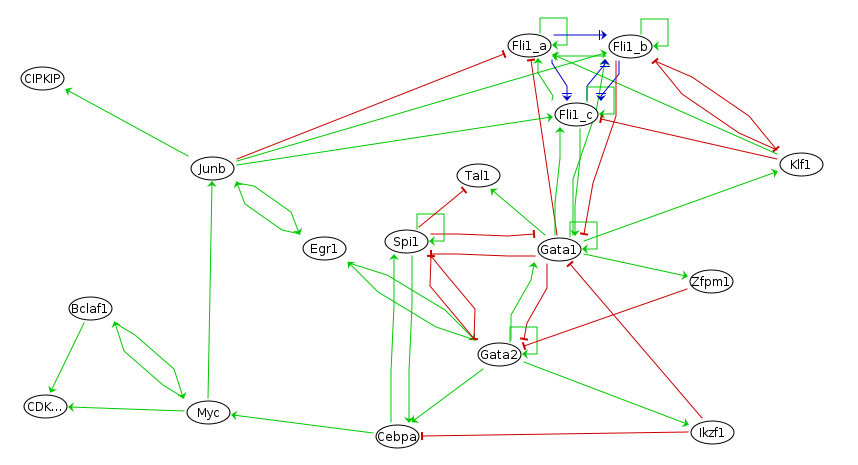}
    \caption{Regulatory graph associated with the model B extended for Fli1.}
    \label{mod_c_et}
\end{figure}

Figure \ref{mod_c_et} allows to visualize the unfolding of the model of Example B for the component Fli1. It features the corresponding triplet of variables, and shows that the relationships with the regulators of Fli1 are consistent with the logical rules described in \ref{regulation}. The rest of the network remains unchanged. With this partial unfolding, the calculation of reachability from an initial configuration becomes possible again. 

\section{Conclusion and outlook}
We have established an unfolding of a Boolean network into another Boolean network that exactly reflects, under asynchronous semantics, the attainability of configurations of the original network under Most Permissive semantics. We implemented a tool performing this unfolding, and extended it to support partial unfolding — restricted to a subset of components — in order to avoid unnecessary model inflation, while preserving the ability to compute reachable states from a given configuration.
Several directions remain open for future work. First, partial unfolding raises the question of the respective roles of individual components in the difference between asynchronous and Most Permissive reachability. A natural step is thus the identification of components that are decisive in these questions of reachability. 

Second, stochastic simulation offers a promising avenue for comparison. Tools such as MaBoSS \cite{maboss} perform stochastic simulation of Boolean networks under asynchronous semantics, and a counterpart for Most Permissive semantics has recently been developed \cite{loic_roncalli}. Comparing the results and computational performance of a simulation run on our unfolded model with MaBoSS against a native Most Permissive simulation would be informative. 

It is therefore possible that the encoding in two steps of the transitions from $i$ to $1$ and from $d$ to $0$ may produce artifacts due to asynchronicity setting.

Finally, Boolean models are intended to support biological research across a range of application domains. Regulatory networks are particularly relevant for modeling cell differentiation processes, with direct implications for cancer research, and the methods developed here contribute to the broader effort of building computationally tractable and biologically meaningful models.

\newpage

\printbibliography

\newpage

\section*{APPENDIX}

\subsection*{ 0 - The Most Permissive dynamics of Example A} 

We give the list of the Most Permissive transitions of Example A. Columns labeled “MP” of Table \ref{MP exA} indicate all these transitions. For instance, “0id —> 010, -di” means that the first $0$ coordinate does not switch, the second becomes $1$ or $d$, the third $0$ or $i$. So 0id has $4$ successors in the Most Permissive STG, that are 01d, 0dd, 0i0 and 0ii. 
\medskip

\begin{table}[h]
\begin{center}
\begin{tabular}{ |l|l|l|l|l |}
\hline 
Synchronous  &  MP             &   MP               &        MP     &         MP                           \\
\hline 
000 —> 001 &  000 —> 00i       &   i00 —> 1ii, d- -   &  d00 —> 0ii, i- -  &    100 —> 1i0               \\
001 —> 001 &  00i —> 001       &   i0i —> 1i1, d-d    &  d0i —> 0i1, i-d   &    10i —> di1, - -d         \\      
010 —> 001 &  00d —> 000, - -i &   i0d —> 1i0, d-i    &  d0d —> 0i0, i-i   &    10d —> di0               \\      
011 —> 001 &  001 —> 001       &   i01 —> 1id, d- -   &  d01 —> 0id        &    101 —> did               \\
100 —> 110 &  0i0 —> 01i, -d-  &   ii0 —> 11i, dd-    &  di0 —> 01i, id-   &    1i0 —> 110               \\
101 —> 010 &  0ii —> 011, -d-  &   iii —> 111, ddd    &  dii —> 011, idd   &    1ii —> d11, - -d         \\
110 —> 110 &  0id —> 010, -di  &   iid —> 110, ddi    &  did —> 010, idi   &    1id —> d10               \\
111 —> 010 &  0i1 —> 011, -d-  &   ii1 —> 11d, dd-    &  di1 —> 01d, -d-   &    1i1 —> d1d               \\
           &  0d0 —> 00i       &   id0 —> 10i, di-    &  dd0 —> 00i, ii-   &    1d0 —> 100, -i-          \\
           &  0di —> 001       &   idi —> 101, did    &  ddi —> 001, iid   &    1di —> d01, -id          \\
           &  0dd —> 000, - -i &   idd —> 100, dii    &  ddd —> 000, iii   &    1dd —> d00, -i-          \\
           &  0d1 —> 001       &   id1 —> 10d, di-    &  dd1 —> 00d, -i-   &    1d1 —> d0d, -i-          \\
           &  010 —> 0di       &   i10 —> 1di, d- -   &  d10 —> 0di, i- -  &    110 —> 110               \\
           &  01i —> 0d1       &   i1i —> 1d1, d-d    &  d1i —> 0d1, i-d   &    11i —> d11               \\
           &  01d —> 0d0, - -i &   i1d —> 1d0, d-i    &  d1d —> 0d0, i-i   &    11d —> d10               \\
           &  011 —> 0d1       &   i11 —> 1dd, d- -   &  d11 —> 0dd        &    111 —> d1d               \\
\hline 
\end{tabular}
\caption{\label{MP exA}Most Permissive transitions for Example A.}
\end{center}
\end{table}

\subsection*{ 1 - Validity of the unfolded model} 
We are going to show that the synchronous function $f_{ext}$ derived from a logical function $f : \mathbb{B}^n\xrightarrow{} \mathbb{B}^n$ in Section \ref{regulation} is indeed a map $f_{ext} : \mathbb{B}^{3n} \xrightarrow{} \mathbb{B}^{3n}$, as expected. Let us recall that, on the other hand, one has in the Most Permissive semantics $f_{m.p.} :X_{m.p.} \xrightarrow[]{} \mathcal{P} (X_{m.p.} ) $, where $X_{m.p.}= \{ 0, i, d, 1 \} ^n$ (cf Remark Section \ref{ss:def-m.p.}). 

\noindent At last, we will prove that the properties of attainability between elements of $\{ 0, i, d, 1 \} ^n $ within the Most Permissive dynamics of $f$ are properly translated by the asynchronous dynamics of $f_{ext}$.

\subsubsection*{The unfolded model as transformation of $\mathbb{B}^{3n}$} In the Most Permissive semantics, we lose the quality of application from $\{ 0, i, d, 1 \}^n$ to itself of the function $f_{m.p.}$, since a configuration $x \in X_{m.p.}$ may be succeeded by two configurations $y$ and $y'$ such that $x_j$, $y_j$ and $y'_{j}$ are all distinct for some value of $j$. This problem may occur with configurations of the form $(x_1, ..., i, ..., x_n) $ or $(x_1, ..., d, ..., x_n) $.
If we focus for instance on configuration $(x_1, ..., i, ..., x_n) $, we see that under suitable conditions the variable previously at $i$ can become $d$ and $1$, changing in two different ways.
In the encoding process, this results in a triplet 001, which can become 011 and 101 (cf Figure \ref{Encodage}). Thus we have no longer the problem of a coordinate that can change in two different ways, since it is not the same coordinate of the triplet that changes. The same goes if we start from a configuration with $d$ instead of $i$. All this guarantees that $f_{ext}$ is an application from $\mathbb{B}^{3n}$ to itself.

\subsubsection*{Preservation of attainability by the unfolded model} 

We have translated the four Most Permissive levels into Boolean triplets, following Table \ref{trad}. In this section we prove that there exists a trajectory between two configurations $x$ and $y$ of $X_{m.p.}$ in the Most Permissive dynamics of $f$ if and only if there exists a trajectory between the corresponding configurations $\tilde{x}$ and $\tilde{y}$ of $\mathbb{B}^{3n}$ (that is according to Table \ref{trad}) in the asynchronous dynamics of $f_{ext}$. 

\begin{table}[h]
\begin{center}
\begin{tabular}{|l|l|}
\hline
$x_j$ & $\tilde{x_j}$ \\ \hline
0    & 000       \\ \hline
i    & 001       \\ \hline
d    & 101       \\ \hline
1    & 111      \\ \hline
\end{tabular}
\caption{\label{trad}Translation of Most Permissive levels into Boolean triplets.}
\end{center}
\end{table}
\begin{itemize}
    \item Let us consider a Most Permissive trajectory $(x^0=x\rightarrow x^1\rightarrow \dots \rightarrow x^r=y)$.
We first translate the sequence of its configurations $(x^k)_{k=0,\dots r}$ in a sequence of configurations $(\tilde{x}^k)_{k=0,\dots r}$ of elements of $\mathbb{B}^{3n}$, using Table \ref{trad}. 
Then, for $k=0, \dots r-1$, the translation of each transition $x^k \rightarrow x^{k+1}$ is made as follows, denoting by $j_k$ the unique index such that $x^k_{j_k} \neq x^{k+1}_{j_k}$: 

\begin{itemize} 
    \item if $(x^k_{j_k},x^{k+1}_{j_k}) \notin \{ (i, 1), (d,0)\}$, then we set one transition $\tilde{x}^k\rightarrow \tilde{x}^{k+1}$,
    \item if $(x^k_{j_k},x^{k+1}_{j_k}) = (i, 1)$, then we set two (asynchronous) transitions $\tilde{x}^k\rightarrow z'\rightarrow \tilde{x}^{k+1}$,\\ the $j$th triplets of $\tilde{x}^k$, $z'$, $\tilde{x}^{k+1}$ being successively $001$, $011$, $111$,
    \item if $(x^k_{j_k},x^{k+1}_{j_k})= (d,0)$, then we set two (asynchronous) transitions $\tilde{x}^k\rightarrow z'\rightarrow \tilde{x}^{k+1}$,\\ the $j$th triplets of $\tilde{x}^k$, $z'$, $\tilde{x}^{k+1}$ being successively $101$, $100$, $000$.
\end{itemize}

In this way we obtain a trajectory from $\tilde{x}$ to $\tilde{y}$.

For instance, in Example A (cf Figures \ref{reg_a}) and \ref{mp_a}), the Most Permissive trajectory\\
$111 \rightarrow d11\rightarrow  dd1 \rightarrow d01 \rightarrow 001 $ is translated that way into \\
$111\;111\;111 \rightarrow101\;111\;111 \rightarrow101\;101\;111 \rightarrow101\;100\;111 \rightarrow101\;000\;111 \rightarrow100\;000\;111 \rightarrow000\;000\;111 $.

However, we must point out that the previous method is not the only way to obtain a trajectory from $\tilde{x}$ to $\tilde{y}$. For instance, the following trajectory is also suitable:

$111\;111\;111 \rightarrow101\;111\;111 \rightarrow101\;101\;111 
\rightarrow101\;100\;111 \rightarrow100\;100\;111 \rightarrow100\;000\;111 \rightarrow000\;000\;111 $.\\
Variants are obtained playing on the positions of transient triplets $100$ or $011$ (the triplet $100$ here).

In conclusion, there is conservation of the attainability between configurations of $X_{m.p.}$ by the unfolded model, but there is no one-to-one correspondence between trajectories linking two elements of $X_{m.p.}$ and trajectories linking their equivalents in the unfolded model.

\item  We still have to show that the unfolded model does not increase attainability. In other words, to show that there is no trajectory $(x'^0=\tilde{x}\rightarrow x'^1\rightarrow \dots \rightarrow x'^r=\tilde{y})$ of the unfolded model, where $\tilde{x}$ and $\tilde{y}$ are associated to elements $x$ and $y$ of $X_{m.p.}$ as above, and such that there exists no Most Permissive trajectory from $x$ to $y$.
Without loss of generality, we suppose that $\tilde{x}$ and $\tilde{y}$ are the only configurations of the trajectory associated to elements of $X_{m.p.}$. This means that all the configurations between $\tilde{x}$ and $\tilde{y}$ involve transient triplets $100$ or $011$. An increase in reachability could then only be due to the presence of these triplets.
However, we have seen in Section \ref{regulation} that the components associated with such triplets can be considered both as active and inactive. In consequence, these triplets authorize the same transitions as the triplets $101$ and $001$ corresponding to $d$ and $i$ they succeed, and therefore cannot contribute to an increase in attainability. 

\end{itemize} 

In conclusion, the attainability between configurations of $X_{m.p.}$ is properly translated by the unfolded model, even if there is no one-to-one correspondence between Most Permissive trajectories and trajectories of the unfolded model.

\subsection*{2 - The function $f_{ext}$ of Example A}

In Example A, with $f_1(x) = x_1 \land  \neg x_3$, $\;f_2(x) = x_1$ and $f_3(x) = \neg x_1$  we get the following regulatory conditions: 
\begin{align*}
\oplus_{1} \;\;&:\; **1 \ *** \ *0* \\
 \neg \oplus_{1} \;\;&:\; **0 \ *** \ *** \lor *** \ *** \ *1*   \\
 \ominus_{1} \;\;&: \;*0* \ *** \ *** \lor *** \ *** \ **1  \\
 \neg \ominus_{1} \;\;&:\; *1* \ *** \ **0 
\end{align*}
  \begin{multicols}{2}
  \begin{align*}
 \oplus_{2} \;\;&:\; **1 \ *** \ ***   \\
 \neg \oplus_{2}  \;\;&:\; **0 \ *** \ ***   \\
 \ominus_{2} \;\;&:\; *0* \ *** \ ***   \\
 \neg \ominus_{2} \;\;&:\; *1* \ *** \ ***   \\
 \end{align*}
 
 \columnbreak
  \begin{align*}
 \oplus_{3} \;\;&:\; *0* \ *** \ ***  \\
 \neg \oplus_{3}  \;\;&:\; *1* \ *** \ ***  \\
 \ominus_{3} \;\;&:\; **1 \ *** \ ***   \\
 \neg \ominus_{3} \;\;&:\; **0 \ *** \ ***  \\
 \end{align*}
\end{multicols}

As a result, le map $f_{ext}$ is given by the following:\begin{align*}
f_{ext}^{-1}(1**\ ***\ ***)&=011\ ***\ ***\;\cup\;110\ ***\ ***\;\cup\;111\ ***\ ***\;\cup\;001\ ***\ ***\;\cup\;101\ *** \ *1* \\
f_{ext}^{-1}(*1*\ ***\ **\ *) &=110\ ***\ ***\;\cup\;0*1\ ***\ ***\;\cup\;111\ ***\ **0 \\
f_{ext}^{-1}(**1\ ***\ ***) &=11*\ ***\ **\ *\;\cup\;0*1\ ***\ *** \\
f_{ext}^{-1}(***\ 1**\ **\ *) &=***\ 011\ **\ *\;\cup\;***\ 110\ **\ *\;\cup\;***\ 111\ **\ *\;\cup\;*0*\ 001\ **\ *\;\cup\;**0\ 101\ **\ * \\
f_{ext}^{-1}(***\ *1*\ ***) &=***\ 110\ **\ *\;\cup\;***\ 0*1\ **\ *\;\cup\;*1*\ 111\ **\ * \\
f_{ext}^{-1}(***\ **\ 1\ **\ *) &=***\ 11*\ ***\;\cup\;***\ 0*1\ **\ *\;\cup\;**1\ 000\ **\ * \\
f_{ext}^{-1}(***\ ***\ 1**) &=***\ ***\ 011\;\cup\;***\ ***\ 110\;\cup\;***\ ***\ 111\;\cup\;**1\ ***\ 001\;\cup\;*1*\ ***\ 101 \\
f_{ext}^{-1}(***\ ***\ *\ 1\ *)&=***\ ***\ 110\;\cup\;***\ ***\ 0*1\;\cup\;**0\ ***\ 111 \\
f_{ext}^{-1}(***\ ***\ **1)  &=***\ ***\ 11*\;\cup\;***\ ***\ 0*1\;\cup\;*0*\ ***\ 000 
\end{align*}

\subsection*{3 - Internal regulations between variables of a triplet} 

\begin{figure}[H]
    \centering

\begin{tikzpicture}

\node (g1) at  (0:2) {$xj\_c$};
\node (g3) at  (144:2) {$xj\_a$};
\node (g2) at  (-120:2) {$xj\_b$};

\draw[->,>=latex, color=blue] (g3) to[bend left=20] (g1);
\draw[->,>=latex, color=blue] (g3) to[bend left=20] (g2);
\draw[->,>=latex, color=green] (g1) to[bend left=8] (g2);
\draw[->,>=latex, color=green] (g1) to[bend left=20] (g3);
\draw[->,>=latex, color=green] (g2) to[bend left=20] (g1);
\draw[->,>=latex, color=green] (g2) to[bend left=20] (g3);
\draw[->,>=latex, color=green] (g3) to[out=150,in=120,looseness=6]  (g3); 
\draw[->,>=latex, color=green] (g1) to[out=30,in=50,looseness=8]  (g1); 
\draw[->,>=latex, color=green] (g2) to[out=-90,in=-120,looseness=14]  (g2);
\end{tikzpicture}

 \caption{\label{reg triplets OK} Guaranteed internal regulations between $x_{ja}$, $x_{jb}$ and $x_{jc}$. Green arrows represent positive regulations. Blue arrows represent regulations with undetermined sign. }
\end{figure}
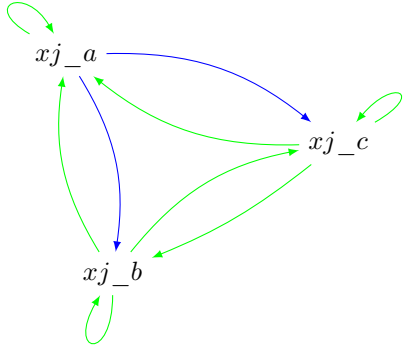

We are going to describe regulations between variables of a triplet, which are directly coming from Figure \ref{Encodage} and internal rules given in Section \ref{internal}.
We recall in Table \ref{triplets} the values the triplets take under the action of $f_{ext}$, following Figure \ref{Encodage}. The character X indicates values that depend actually on the index $j$ of the component represented by the triplet, and on the considered configuration of $\mathbb{B}^{3n}$, which may follow or not the given additional condition of the internal rules ($\oplus_j$, $\neg \oplus_j$, $\ominus_j$ or $\neg \ominus_j$). From Table \ref{triplets} we can deduce immediately the guaranteed internal regulations between the variables $x_{ja}$, $x_{jb}$ and $x_{jc}$, represented in Figure \ref{reg triplets OK}.

\begin{table}[h]
\begin{center}
\begin{tabular}{|l|l|}
\hline
0 0 0     & 0 0 X \\ \hline
0 0 1     & X 1 1       \\ \hline
0 1 0    & 0 0 0       \\ \hline
0 1 1    & 1 1 1       \\ \hline
1 0 0    & 0 0 0      \\ \hline
1 0 1    & X 0 0      \\ \hline
1 1 0    & 1 1 1      \\ \hline
1 1 1    & 1 X 1      \\ \hline

\end{tabular}
\caption{\label{triplets}Evolution of the triplets under $f_{ext}$ (synchronous updating).}
\end{center}
\end{table}

Three other regulations may occur, not already seen above as guaranteed internal regulations, and that are depending on conditions.

\begin{itemize}
    \item Inhibition of $ x_{jc}$ by $x_{jb}$:
    
    Given $j\in\{1,\dots,n\}$ and a configuration $x\in \mathbb{B}^{3n}$ whose $j$th triplet is equal to $000$, this triplet becomes $001$ under $f_{ext}$ if and only if the condition $\oplus_j$ is fulfilled by $x$. Moreover, triplet $010$ becomes $000$ under $f_{ext}$. 
        So there is an inhibition of $ x_{jc}$ by $x_{jb}$ if and only if there exists such a configuration $x$.

        In Example A, there is no inhibition of $ x_{1c}$ by $x_{1b}$, given that $\oplus_{1} : **1 \ *** \ *0*$. The expression $ \oplus_{2} : **1 \ *** \ *** $ 
        shows that there is an inhibition of $ x_{2c}$ by $x_{2b}$, via configurations of $**1 \ 000 \ ***$. Likewise the expression $\oplus_{3} : *0* \ *** \ *** $ 
        shows an inhibition of $ x_{3c}$ by $x_{3b}$, via configurations of $*0* \ *** \ 000$.

          \item Inhibition of $ x_{jb}$ by $x_{jc}$:
          
          Given $j\in\{1,\dots,n\}$ and a configuration $x\in \mathbb{B}^{3n}$ whose $j$th triplet is equal to $111$, this triplet becomes $101$ under $f_{ext}$ if and only if the condition $\ominus_j$ is fulfilled by $x$. Moreover, triplet $110$ becomes $111$ under $f_{ext}$. 
          
So there is an inhibition of $ x_{jb}$ by $x_{jc}$ if and only if there exists such a configuration $x$.

          As above, in Example A, the expressions of $\ominus_{1} :*0* \ *** \ *** \lor *** \ *** \ **1$, $ \ominus_{2} : *0* \ *** \ ***  $ and $ \ominus_{3} : **1 \ *** \ ***  $ lead to the three inhibitions of $ x_{1b}$ by $x_{1c}$, of $ x_{2b}$ by $x_{2c}$ and of $ x_{3b}$ by $x_{3c}$.

    \item Self-inhibition of $ x_{ja}$:
    
    Given $j\in\{1,\dots,n\}$ and a configuration $x\in \mathbb{B}^{3n}$ whose $j$th triplet is equal to $001$, this triplet becomes $111$ under (the synchronous action of) $f_{ext}$ if and only if the condition $\ominus_j$ is fulfilled by $x$.

            Given a configuration $y\in \mathbb{B}^{3n}$ whose $j$th triplet is equal to $101$, this triplet becomes $000$ under (the synchronous action of) $f_{ext}$ if and only if the condition $ \oplus_j$ is fulfilled by $y$. 
            
        If these two configurations $x$ and $y$ may differ only by the first coordinate of the $j$th triplet, there is a self-inhibition of $ x_{ja}$.
       
        In Example A, given fixed elements $k,\ p,\ q,\ r, \ s,\ t,\ u$ of $\{0,1\}$, 

        - considering $x=001\ pqr\ s0u$ and $y= 101\ pqr\ s0u$ shows a self-inhibition of  $x_{1a}$,

         - considering $x=k01\ 001\ stu$ and $y= k01\ 101\ stu$ shows a self-inhibition of  $x_{2a}$,

          - considering $x=k01\ pqr\ 001$ and $y= k01\ pqr\ 101$ shows a self-inhibition of  $x_{3a}$.
          
\end{itemize}

\subsection*{4 - About the regulatory graph of $f_{ext}$ }

On the basis of example A, we have compared the interactions of the regulatory graph $RG(f_{ext})$ of $f_{ext}$ with the interactions of the regulatory graph $RG(f)$ of $f$, considering all the pairs of states coding Most Permissive states where they are expressed. Our attention was focused in particular on reversed signs of interactions, and on interactions possibly produced by the Most Permissive dynamics but not by the asynchronous one.
\bigskip

- The $RG(f_{ext})$ interactions involving two distinct triplets all reflect $RG(f)$ interactions. Their signs are translated directly, or they are reversed in consequence of the complexity of their transcription.\\ 
For example, $x_1$ activates $x_2$ in $RG(f)$, and $x_{1c}$ activates $x_{2c}$ in $RG(f_{ext})$. This activation is expressed for instance at the pair of states $\{000\ 000 \ 000$, $001\ 000 \ 000\}$ through transitions 00\textbf{0} 000 000 --> 000 00\textbf{0} 001 and 00\textbf{1} 000 000 --> 001 00\textbf{1} 000, encoding the Most Permissive transitions 000 --> 00i and i00 --> ii0. The activation of $x_2$ by $x_1$ is thus clearly translated.\\
In another example, $x_3$ inhibits and does not activate $x_1$ in $RG(f)$, when $x_{3b}$ activates $x_{1a}$ in $RG(f_{ext})$. This activation is expressed for instance at the pair of states $\{101\ 101 \ 101$, $101\ 101 \ 111\}$ through transitions 101 000 1\textbf{0}1 --> \textbf{0}01 000 101 and 101 000 1\textbf{1}1 --> \textbf{1}00 000 111 --> 000 000 111. Thus this pair of states shows locally an activation of $x_{1a}$ by $x_{3b}$, but this activation is linked to the construction of the coding: there is no interpretation to give in terms of activation on the Most Permissive side. The only interpretation is that these trajectories reflect the inhibition of $x_1$ by par $x_3$ of the initial model, through the encoded transitions d0d --> i0d and d01 --> 001.

\medskip
- For the $RG(f_{ext})$ interactions involving one single triplet, we did not consider guaranteed regulations.
Among the non-guaranteed regulations, we only took into account the self-inhibitions of $ x_{1a}$, $ x_{2a}$ and $ x_{3a}$, the other ones involving artifactual or intermediate triplets. These three self-inhibitions reflect three Most Permissive self-inhibitions. Remark that $x_2$ and $x_3$ are not self-inhibited in $RG(f)$, so that we may consider that these new self-inhibitions are produced by the Most Permissive dynamics, and not by the asynchronous one. Indeed, these inhibitions are produced by the frequent flip-flop i <---> d in the Most Permissive dynamics.\\ 
For example, the self-inhibition of $ x_{3a}$ is expressed at the pair of states $\{001\ 000 \ 001$, $001\ 000 \ 101\}$ through transitions 
001 000 \textbf{0}01 --> 001 000 \textbf{1}01 and 001 000 \textbf{1}01 --> 001 000 \textbf{0}01, encoding Most Permissive transitions i0i --> i0d and i0d --> i0i, which do translate a self-inhibition.

In conclusion, based on Example A, Most Permissive semantics does not seem to create any meaningful new interactions.
This is in line with the fact that it essentially encodes refinements of the starting model, which do not change the structure of the regulation graph.

\end{document}